\documentclass[journal,twoside]{IEEEtran}
\usepackage{makecell}
\usepackage{amsfonts}
\usepackage{graphicx}
\usepackage{amsmath}
\usepackage{subfigure}
\usepackage{multirow}
\usepackage{booktabs}
\usepackage{algorithm}
\usepackage{CJK}
\usepackage{amsthm}
\usepackage{bm}
\usepackage[square, comma, sort&compress, numbers]{natbib}
\usepackage{hyperref}
\usepackage{amssymb}
\usepackage{algpseudocode}
\usepackage{array}
\usepackage[table]{xcolor}

\begin{document}
\newtheorem{myDef}{Definition}
\renewcommand{\algorithmicrequire}{\textbf{Input:}} 
\renewcommand{\algorithmicensure}{\textbf{Output:}} 
\newcommand{\minitab}[2][l]{\begin{tabular}{#1}#2\end{tabular}}

\title{Efficient Sparse Code Multiple Access Decoder Based on Deterministic Message Passing Algorithm}

\author{Chuan~Zhang,~\IEEEmembership{Member,~IEEE},
        Chao~Yang,~
        Wei~Xu,~\IEEEmembership{Senior Member,~IEEE},
        Shunqing~Zhang,~\IEEEmembership{Senior Member,~IEEE},
        Zaichen~Zhang,~\IEEEmembership{Senior Member,~IEEE},
        and~Xiaohu~You,~\IEEEmembership{Fellow,~IEEE}

\thanks{Chuan Zhang and Chao Yang are with Lab of Efficient Architectures for Digital-communication and Signal-processing (LEADS), Southeast University, Nanjing, China.
Chuan Zhang, Chao Yang, Wei Xu, Zaichen Zhang, and Xiaohu You are with the National Mobile Communications Research Laboratory, Southeast University, Nanjing, China.
Chuan Zhang, Chao Yang, and Zaichen Zhang are with Quantum Information Center of Southeast University, Nanjing, China. Email: \{chzhang, chaoyang, wxu, zczhang, xhyu\}@seu.edu.cn.}
\thanks{Shunqing Zhang is with Shanghai Institute for Advanced Communications and Data Science, Shanghai University, Shanghai, China. Email: shunqing@shu.edu.cn.}
\thanks{This paper was presented in part at IEEE Asia Pacific Conference on Circuits and Systems (APCCAS), Jeju, Korea, 2016, as a Best Paper Award recipient. Chuan Zhang and Chao Yang contributed equally to this work. \emph{(Corresponding author: Chuan Zhang.)}}
}

\markboth{IEEE Transactions on ,~2018}%
{C. Zhang \MakeLowercase{\textit{et al.}}: Efficient Sparse Code Multiple Access Decoder Based on Deterministic Message Passing Algorithm}

\maketitle

\begin{abstract}
Being an effective non-orthogonal multiple access (NOMA) technique, sparse code multiple access (SCMA) is promising for future wireless communication. Compared with orthogonal techniques, SCMA enjoys higher overloading tolerance and lower complexity because of its sparsity. In this paper, based on deterministic message passing algorithm (DMPA), algorithmic simplifications such as domain changing and probability approximation are applied for SCMA decoding. Early termination, adaptive decoding, and initial noise reduction are also employed for faster convergence and better performance. Numerical results show that the proposed optimizations benefit both decoding complexity and speed. Furthermore, efficient hardware architectures based on folding and retiming are proposed. VLSI implementation is also given in this paper. Comparison with the state-of-the-art have shown the proposed decoder's advantages in both latency and throughput (multi-Gbps).
\end{abstract}


\IEEEpeerreviewmaketitle

\section{Introduction}

\IEEEPARstart{T}{he} fifth generation of cellular network (5G) is put forward to meet the ever-increasing demand of wireless communication. Enabling techniques of 5G include massive multiple-input multiple-output (MIMO), advanced coding, new multiple access (MA), full spectrum access, new network architectures, etc \cite{5g_tech}. In the past decades, MAs such as time division multiple access (TDMA) \cite{TDMA}, frequency division multiple access (FDMA) \cite{FDMA}, and code division multiple access (CDMA) \cite{CDMA}, became part of wireless standards. However, those orthogonal MAs can hardly meet the 5G's capacity requirement ($10^3$ times of LTE), due to limitations on multiplexing approaches towards physical resources \cite{5g}. According to 3GPP white book, in the enhanced Mobile Broadband (eMBB) scenario, the peak data rate should be $20$ Gbps ($10$ to $10^2$ times of LTE), the peak spectral efficiency should be $30$ bps/Hz ($3$ to $5$ times of LTE), and the latency should be less than $1$ ms ($10\%$ of LTE) \cite{3gpp,3gpp_time}. Thus, ideas of non-orthogonal MA (NOMA) \cite{NOMA_tech} are proposed to alleviate these bottlenecks.

\subsection{Challenges for Existing NOMA}
Compared to orthogonal MAs, NOMA techniques refer to those allowing multiple users overlap in time, frequency, or code domain, in other words, sharing the same physical resources \cite{NOMA_intro}. NOMA is able to distinguish different users via successive interference cancellation (SIC) \cite{SIC} or multiple user decoding (MUD) \cite{MUD}. Besides the very first version \cite{NOMA_detail}, the state-of-the-art (SOA) NOMA includes multiuser shared access (MUSA) \cite{MUSA_detail}, pattern division multiple access (PDMA) \cite{PDMA_detail}, sparse code multiple access (SCMA) \cite{nikopour2013sparse}, etc. SIC was employed in \cite{NOMA_detail,MUSA_detail,PDMA_detail} and has practical challenges:
\begin{itemize}
  \item Computational complexity: SIC implies that each user can be decoded only when all the prior users are properly decoded. Therefore, its computational complexity scales with the in-cell user number.
  \item Error propagation: For SIC, if an error occurs, all users afterward are likely to be decoded incorrectly.
  \item Decoding latency: User power sorting is involved in SIC, and causes good overhead latency compared to other methods. Since the data with the lowest power is decoded last, the latency will even higher.
\end{itemize}
Therefore, SCMA employs MUD instead of SIC. Thanks to its sparsity, message passing algorithm (MPA) can be applied for better decoding performance.

\subsection{Sparse Code Multiple Access}
SCMA was proposed in 2013 \cite{nikopour2013sparse}, trying to increase user scale via a new perspective: enabling more efficient multiple access by non-orthogonal sparse spreading codes of users.

\subsubsection{Properties of SCMA}
As a promising MA, SCMA has the properties: i) multiplexing in frequency domain; ii) codebook based on both mapping and spreading; iii) multi-dimensional constellation for shaping gain and spectral efficiency; iv) non-orthogonality ensuring more accessed users; v) spreading which reduces noise interference and enhances system robustness; and vi) sparsity which reduces decoding complexity. Thanks to these properties, SCMA is more physically realizable and overloading tolerant, compared to other MAs \cite{nikopour2014scma}. Details of SCMA can be found in Section \ref{sec:Preliminaries}.

\subsubsection{Challenges of SCMA}
\begin{itemize}
  \item Throughput: Though the throughput of SCMA outperforms other MAs, especially orthogonal ones, it is hard to achieve the eMBB peak rate with acceptable complexity. Admittedly, such throughput can be achieved with a larger overloading factor, leading to prohibitive hardware complexity and performance degradation.
  \item Latency: On one hand, utilizing MUD, SCMA avoids the sorting latency required by SIC. On the other hand, for imperfect channels the iterative MPA tends to cost more iterations, which will counteract its latency advantage.
  \item Implementation: Though VLSI techniques ensure that complexity is no longer a bottleneck for SCMA implementation when the overloading factor is not extremely large, existing iterative algorithms are not hardware friendly. Second, the noise power density $N_0$ results in large data range, leading to unbearable quantization length, or otherwise poor error performance.
\end{itemize}

\subsection{Relevant Prior Art}
Regarding SCMA decoding, existing literature mainly focus on three aspects: i) stochastic computing, ii) tree structure approximation, and iii) efficient hardware implementation.

\subsubsection{Stochastic Computing}
In \cite{SMPA_1}, a stochastic MPA (SMPA) decoder was proposed, where beliefs are given by weights of bit streams. Multiplication and addition are implemented by \textsc{and} and \textsc{mux}, respectively. Though it work effectively reduces the complexity per iteration, problems are:

\begin{itemize}
\item Accuracy: Stochastic computing suffers from low accuracy, due to randomness loss. Beliefs in MPA usually require precision of $10^{-5}$, which length-limited could not give. Performance degradation is observed.
\item Latency: For SMPA, the calculation of a single value requires a large number ($10^{5}$ to $10^{6}$) of bit-level operations. Considerable iterations make the latency even larger and not suitable for practice.
\item Complexity: Though SMPA helps to reduce hardware of a single operation, the amount of bit-operations in one decoding is around $10^{7}$. Thus, the total complexity may be even larger than deterministic MPA (DMPA).
\end{itemize}
A VLSI architecture of SMPA was discussed in \cite{SMPA_2}. The throughput for a $6$-user decoder is $57$ Mbps and far from 3GPP requirements. Though the hardware cost is low, the latency is not suitable for eMBB.

\subsubsection{Tree Structure Approximation}
In \cite{tree_1}, a pruned tree approximation was proposed.
The decoder accurately represents values with high probabilities, whereas approximates ones with low probabilities \cite{fossorier1999reduced}. Squares are replaced by additions, multiplications, and comparisons. Though complexity is expected to reduce, search breadth must be larger than $2$ for performance, which increases the complexity again.

\subsubsection{Efficient Hardware Architecture}
In \cite{SCMA_archi}, a stage-level folded architecture for DMPA was proposed with consideration of both speed and efficiency, which is our prior work. However, only theoretical analysis and simple architecture were given. The real VLSI implementation is missing.

\subsection{Contributions}
This paper emphasises on iteration reduction, convergence speedup, computation simplification, and implementation of SCMA decoder. Compared to SOA, our contributions are:
\begin{itemize}
  \item We propose early termination scheme based on the convergence behavior of DMPA, which significantly reduces the required iteration number.
  \item We propose adaptive decoder, which adjusts beliefs according to the variation trend, accelerates the convergence, and compensates the performance loss. Results show that it outperforms the ones in \cite{SMPA_1,SMPA_2} in terms of latency and throughput, satisfying the 3GPP requirements.
  \item We perform numerical analysis for conditional probability approximation (over $60\%$ computation is for conditional probabilities in MPA) in \emph{Initialization}, which is square-free and division-free, and suffers from little performance loss. Computational complexity and hardware implementation have been greatly benefitted.
  \item We propose distributed matrix scheme for prior noise reduction of DMPA decoder, which compensates the approximation loss with negligible extra complexity.
  \item We improve our stage-level folded decoder with the proposed algorithms, achieve higher hardware efficiency with eMBB requirements on throughput and latency.
  \item We implement the proposed DMPA decoder on Xilinx Virtex-7 XC7VX690T FPGA to demonstrate its advantages for real applications.
\end{itemize}


\subsection{Notations}
Lowercase and uppercase boldface letters designate column vectors and matrices, respectively. Matrix $\bf{A}$'s transpose and conjugate are ${\bf{A}}^{T}$ and ${\bf{A}}^{H}$. The $M \times M$ identity matrix is ${\bf{I}}_{M}$ and the $M \times N$ all-zeros matrix is ${\bf{0}}_{M \times N}$. Sets are denoted by uppercase calligraphic letters $\mathcal{A}$, with cardinality $|\mathcal{A}|$.

\subsection{Paper Outline}
The remainder of this paper is organized as follows. Section \ref{sec:Preliminaries} reviews the preliminaries of SCMA. DMPA and its optimized versions are discussed in Section \ref{sec:Optimized}. Numerical results and analysis are given in Section \ref{sec:Results}. Hardware architecture is described in Section \ref{sec:Hardware}. VLSI implementation is given in Section \ref{sec:VLSI}. Section \ref{sec:Conclusion} concludes the entire paper.

\section{Preliminaries}\label{sec:Preliminaries}
\begin{figure}[htbp]
\centering
\includegraphics[width=.9\linewidth]{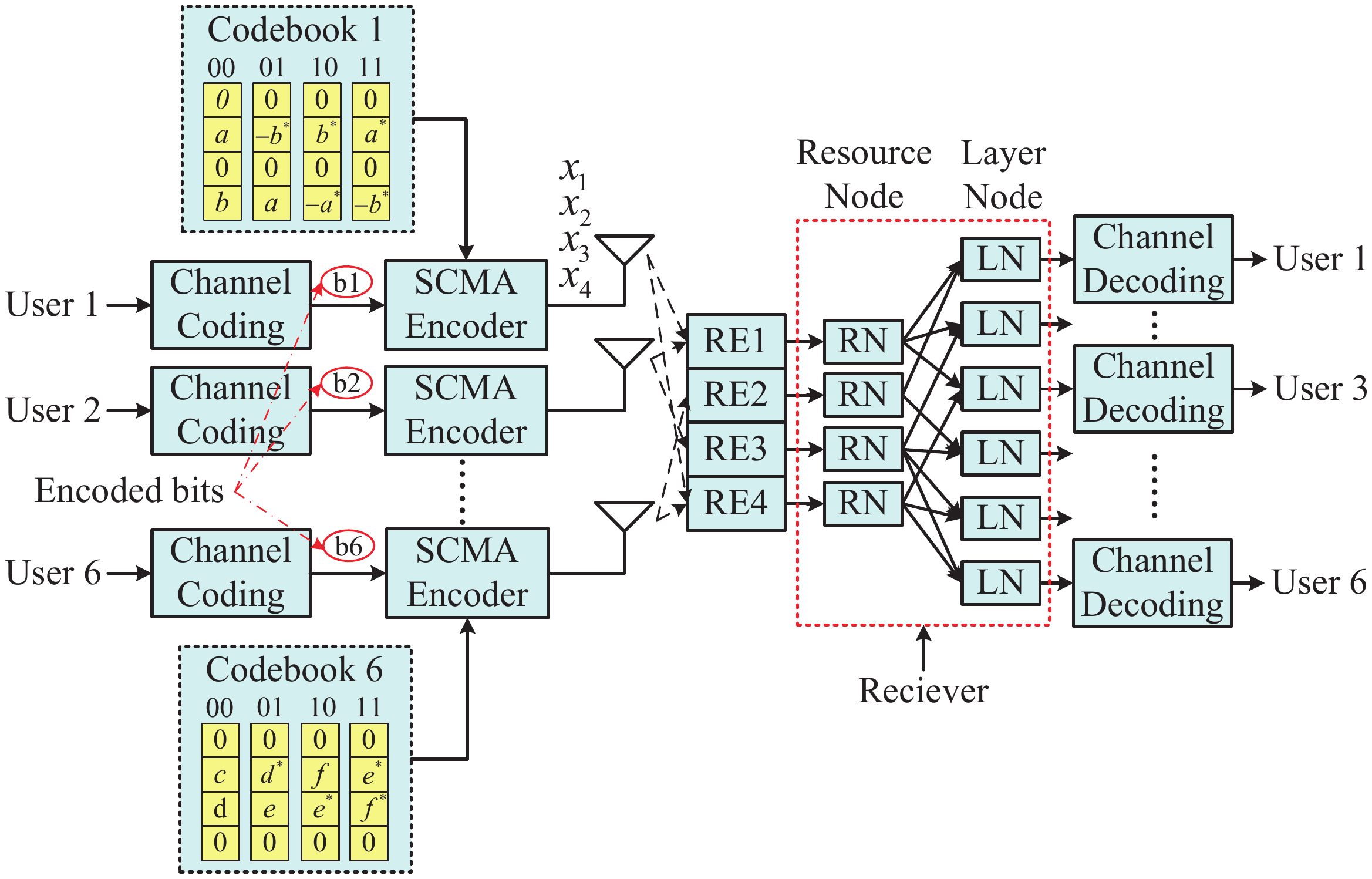}
\caption{\label{fig:SCMA_sys}A $6$-user SCMA system.}
\end{figure}

Preliminaries of SCMA are given in this section. A $6$-user system in Fig. \ref{fig:SCMA_sys} is used as a running example.

\subsection{SCMA Encoder}
Suppose codeword set, constellation set, and information set are $\mathcal{X}$, $\mathcal{C}$, and $\mathcal{B}$, respectively. Define ${\bf{x}} \in \mathcal{X}$, ${\bf{c}} \in \mathcal{C}$, and ${\bf{b}} \in \mathcal{B}$. $|\mathcal{B}|=M$, $|{\mathcal{X}}|=K$, and $|{\mathcal{C}}|=N$. The SCMA encoding is given by two rounds of mapping \cite{nikopour2013sparse}. The first round of mapping is:
\begin{equation}
g:~\mathcal{B} \rightarrow \mathcal{C},~{\bf{c}} = g ({\bf{b}}),
\end{equation}
where $\mathcal{B} \subset \mathbb{B}^{\log_{2}M}$, $\mathcal{C} \subset \mathbb{C}^{N}$, and $g$ is a constellation mapping function. The second round of mapping is:
\begin{equation}
{\bf{V}}:~\mathcal{C} \rightarrow \mathcal{X},~{\bf{x}} = {\bf{Vc}},
\end{equation}
where $\mathcal{X} \subset \mathbb{C}^{K}$, and ${\bf{V}} \in \mathbb{B}^{K \times N}$ is the mapping matrix.

Suppose the entire mapping function of SCMA encoding is $f$. Then we have
\begin{equation}
f:~\mathcal{B} \rightarrow \mathcal{X},~{\bf{x}} = f ({\bf{b}}),~f = {\bf{V}}g.
\end{equation}
An $M$-size SCMA codebook consisting of $K$ complex values is constructed.
Note that ${\bf{V}}$ contains $(K-N)$ all-zero rows. Mapping matrix is generated by inserting $(K-N)$ all-zero rows into an $N \times N$ identity matrix ${\bf{I}}_{N}$ randomly. So when the SCMA system is regular, it supports $C^{N}_{K} = C^{K-N}_{K}$ different layers (users).

\subsection{SCMA Multiplexing}
Consider a $K$-dimensional SCMA encoder with $J$ separated layers. Each layer is defined by $({\bf{V}}_j, {\bf{g}}_j, {\bf{M}}_j, {\bf{N}}_j)$, where $j = 1,...,J$. If $i \neq j$, ${\bf{V}}_i \neq {\bf{V}}_j$ and ${\bf{g}}_i \neq {\bf{g}}_j$, in order to distinguish one layer from another. In general, ${\bf{M}}_j$ and ${\bf{N}}_j$ can be either the same or different for different layers. Without loss of generality, for $\forall j$ we set ${\bf{M}}_j = M$, ${\bf{N}}_j = N$.

We call this SCMA system semi-regular because $J$ is not necessarily $C^{N}_{K}$ (The regular system will be discussed later). The SCMA codewords are multiplexed over $K$ shared orthogonal resources, e.g. OFDMA tones or MIMO spatial layers \cite{nikopour2014scma}. With this semi-regular system, the received signal after synchronous layer multiplexing can be expressed as
\begin{equation}\label{scma1}
{\bf{y}} = \Sigma^{J}_{j=1}diag({\bf{h}}_{j}){\bf{x}}_{j} + {\bf{n}},
\end{equation}
where ${\bf{h}}_{j}$ and ${\bf{x}}_{j}$ are the $K$-dimensional channel vector and SCMA codeword of layer $j$. Suppose signals of all layers are from the same transmit point, for a specific receiver, the channel vectors of all layers are identical that for $\forall j$, ${\bf{h}}_j = {\bf{h}}$. Now Eq. (\ref{scma1}) reduces to
\begin{equation}
{\bf{y}} = diag({\bf{h}})\Sigma^{J}_{j=1}{\bf{x}}_{j} + {\bf{n}}.
\end{equation}

\begin{figure}[htbp]
\centering
\includegraphics[width=.9\linewidth]{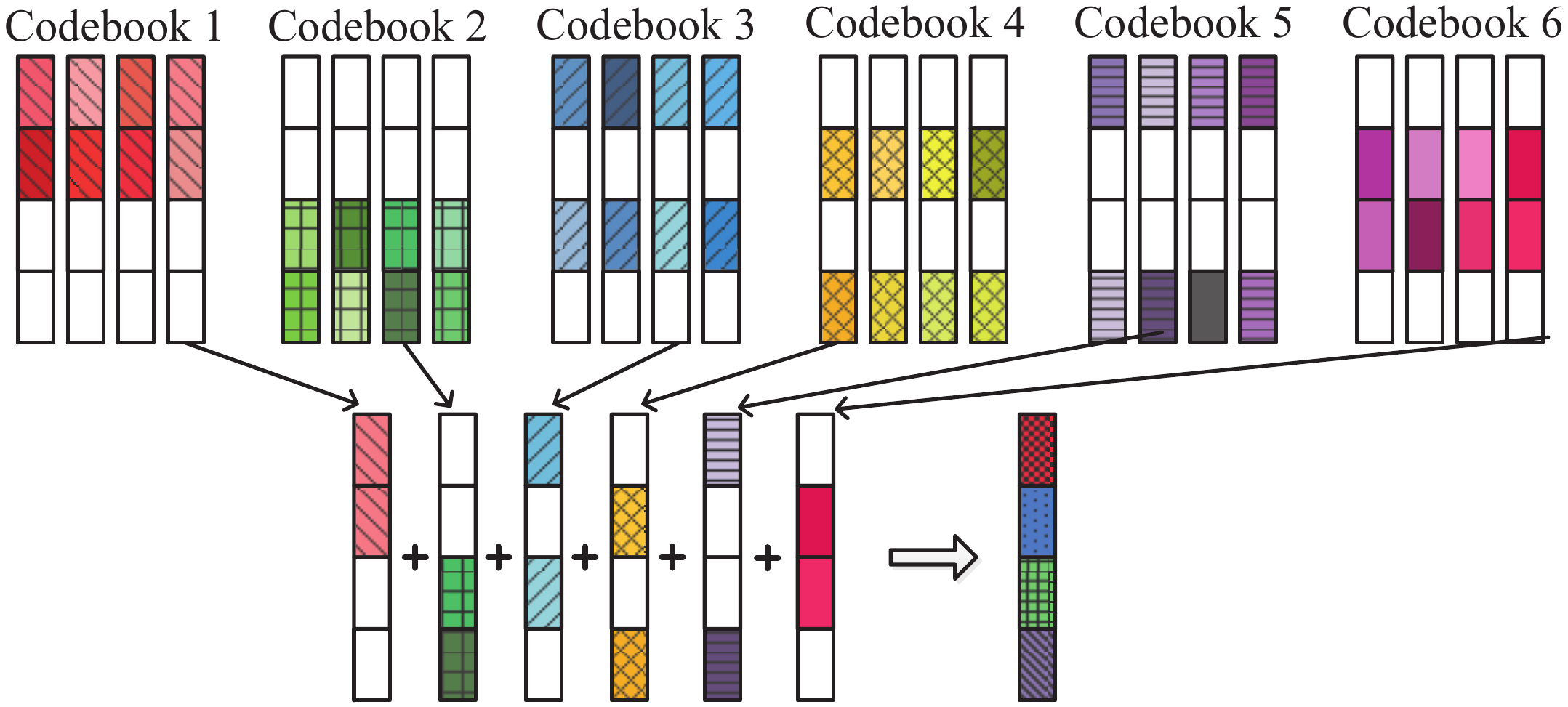}
\caption{\label{fig:multiplex}SCMA multiplexing example.}
\end{figure}

Define overloading factor as $\lambda = J/K$, which indicates the overloading tolerance or access ability of a SCMA system. Fig. \ref{fig:multiplex} illustrates a $6$-user SCMA multiplexing.

\subsection{Factor Graph Representation}
Define the binary indicator vector as ${\bf{f}}_j = diag({\bf{V}}_j{\bf{V}}^{T}_{j})$. Then the factor graph matrix is ${\bf{F}} = ({\bf{f}}_1,...,{\bf{f}}_J)$. Then the factor graph representation can be obtained like how we do with LDPC codes. Each column of ${\bf{F}}$ associates a layer node, and each row a resource node. Degree of each resource node is defined as ${\bf{d}}_{f} = (d_{f1},...,d_{fK})^{T} = \Sigma^{J}_{j=1}{\bf{f}}_j$. For more details, please refer to \cite{nikopour2013sparse}.

\begin{figure}[htbp]
\centering
\includegraphics[width=.75\linewidth]{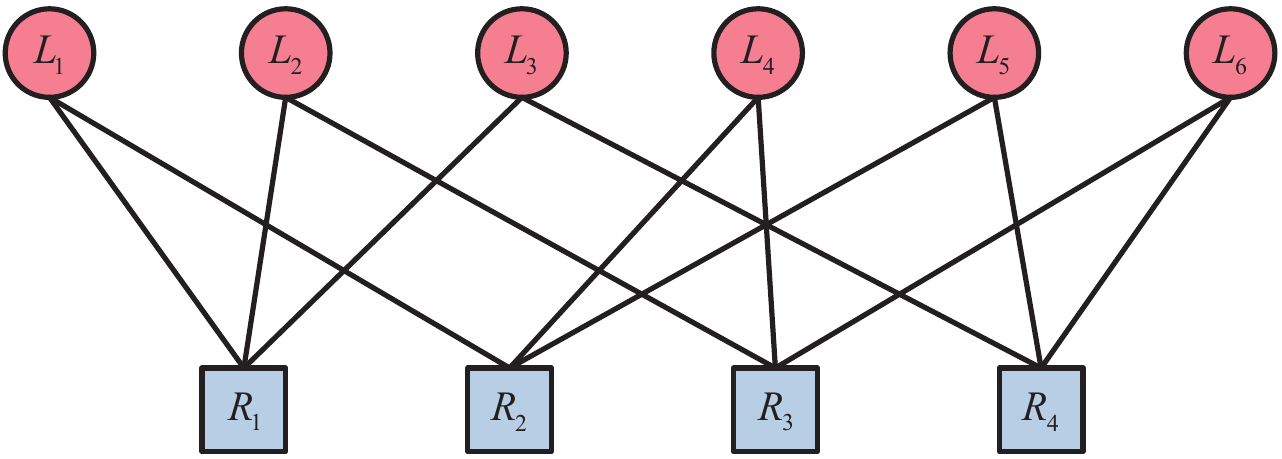}
\caption{\label{fig:fac_graph}Factor graph representation of an SCMA with $K = 4$ and $J = 6$. }
\end{figure}

Take $K = 4$ and $N = 2$ as an example. The factor graph is in Fig. \ref{fig:fac_graph} and $J = C^{2}_{4} = 6$. Degree ${\bf{d}}_{f} = (d_{f1},...,d_{fK})^{T} = (3,3,3,3,3,3)^{T}$ and the overloading factor $\lambda = J/K = 1.5$. The $4\times6$ factor graph matrix of this system is in Eq. (\ref{matrix}).
\begin{gather}\label{matrix}
{\bf{F}} =
\begin{bmatrix}
1 & 1 & 1 & 0 & 0 & 0\\
1 & 0 & 0 & 1 & 1 & 0\\
0 & 1 & 0 & 1 & 0 & 1\\
0 & 0 & 1 & 0 & 1 & 1
\end{bmatrix}
\end{gather}

\section{Optimizations on SCMA Decoding}\label{sec:Optimized}
%
\subsection{Regular Form of SCMA}
Regular SCMA refers to the absolute-regular form \cite{ryan2002low,gallager1962low}, where number of layers $J$ equals to $C^{N}_{K}$. In other words, it employs all the available layers (users). Eq. (\ref{matrix}) is an example of regular form. The definition is as follows.
\begin{myDef}
SCMA with $K$ complex-dimension and weight of $N$ which satisfies the following requirements is called regular (absolute-regular) SCMA.\\
Requirement $1$: Owning $J = C^{N}_{K}$ layers (users) in total.\\
Requirement $2$: The columns of factor graph matrix must be listed in the sequential permutation order, with weight $\lambda$.
\end{myDef}
%
\subsection{DMPA Decoding}
The DMPA decoding for SCMA mainly includes $4$ steps.

\subsubsection{Initialization}
Calculate conditional probability with extrinsic information to get prepared for the belief propagation.
\begin{equation}
\resizebox{.91\linewidth}{!}{$
\label{eqn:2}
P_{k}(y_{k}|x_{k,1},x_{k,2},x_{k,3},N_{0}) = e^{-\parallel y_{k}-(x_{k,1}+x_{k,2}+x_{k,3})\parallel^{2}/N_{0}},
$}
\end{equation}
where $y_{k}$ denotes the $k$-th bit of the received signal ${\bf{y}}$. $x_{k,1}$, $x_{k,2}$, and $x_{k,3}$ denote overlapped bits of the $3$ layers which are connected to the $k$-th resource node separately, and $N_{0}$ is the noise power density.

\subsubsection{Resource Node Updating}
The updating formulation of resource node is in the sum-product form which is an approximation of marginal probability:
\begin{equation}
\label{eqn:3}
\resizebox{.91\linewidth}{!}{$
I_{R_{k}\rightarrow L_{1}}(m_{1}) = \sum^{M}_{m_{2}=1}\sum^{M}_{m_{3}=1}P_{k}I_{L_{2}\rightarrow R_{k}}(m_{2})I_{L_{3}\rightarrow R_{k}}(m_{3}),
$}
\end{equation}
\begin{equation}
\label{eqn:4}
\resizebox{.91\linewidth}{!}{$
I_{R_{k}\rightarrow L_{2}}(m_{2}) = \sum^{M}_{m_{1}=1}\sum^{M}_{m_{3}=1}P_{k}I_{L_{1}\rightarrow R_{k}}(m_{1})I_{L_{3}\rightarrow R_{k}}(m_{3}),
$}
\end{equation}
\begin{equation}
\label{eqn:5}
\resizebox{.89\linewidth}{!}{$
I_{R_{k}\rightarrow L_{3}}(m_{3}) = \sum^{M}_{m_{1}=1}\sum^{M}_{m_{2}=1}P_{k}I_{L_{1}\rightarrow R_{k}}(m_{1})I_{L_{2}\rightarrow R_{k}}(m_{2}),
$}
\end{equation}
where $R_{k}$ is the $k$-th resource node, $m_{1,2,3} = 1,...,M$ are transmitted symbols. $I_{R_{k}\rightarrow L_{1,2,3}}$ denotes the belief propagated to the $k$-th resource node from the neighboring layer nodes. $I_{L_{1,2,3}\rightarrow R_{k}}$ is the belief passing in the opposite direction.

\subsubsection{Layer Node Updating}
The normalization makes sure belief falls in $[0,1]$.
\begin{equation}
\label{eqn:6}
I_{L_{j}\rightarrow R_{1}}(m) = normalize(I_{R_{2}\rightarrow L_{j}}(m)),
\end{equation}
\begin{equation}
\label{eqn:7}
I_{L_{j}\rightarrow R_{2}}(m) = normalize(I_{R_{1}\rightarrow L_{j}}(m)),
\end{equation}
where $m = 1,...,M$ corresponds different symbols.

\subsubsection{Probability Calculating and Symbol Judging}
After iterations, the final probability of each symbol is
\begin{equation}
\label{eqn:8}
Q_{L_{j}}(m) = I_{R_{1}\rightarrow L_{j}}(m)\cdot I_{R_{2}\rightarrow L_{j}}(m).
\end{equation}
where $L_{j}$ denotes the $j$-th layer. The symbol with the highest probability becomes the estimated symbol $\bm{\hat{l}}$ for each layer.

\subsection{Max-Log Algorithm}
Decoder in probability domain suffers from huge complexity and relatively high latency. Therefore, its Max-Log version is considered \cite{chen2005reduced} with the Jacobi's logarithm formula \cite{robertson1995comparison}:
\begin{equation}
\label{eqn:jacobi}
\log \Bigg( \sum_{i=1}^{N}\mathrm{exp}(f_{i}) \Bigg) \approx \max_{i=1,...,N}\{f_{1},f_{2},...,f_{N}\}.
\end{equation}
Updating steps now become:
\subsubsection{Initialization}
\begin{equation}
\label{equ:log1}
\resizebox{.89\linewidth}{!}{$
P_{k}^{\log}(y_{k}|x_{k,1},x_{k,2},x_{k,3},N_{0}) = -\frac{1}{N_{0}}\parallel y_{k}-(x_{k,1}+x_{k,2}+x_{k,3})\parallel^{2},
$}
\end{equation}
\subsubsection{Resource Node Updating}
\begin{equation}
\label{eqn:log2}
\resizebox{.89\linewidth}{!}{$
I_{R_{k}\rightarrow L_{1}}^{\log}(m_{1}) = \max \Big\{P_{k}^{\log}+I_{L_{2}\rightarrow R_{k}}^{\log}(m_{2})+I_{L_{3}\rightarrow R_{k}}^{\log}(m_{3})\Big\},
$}
\end{equation}
\begin{equation}
\label{eqn:log3}
\resizebox{.89\linewidth}{!}{$
I_{R_{k}\rightarrow L_{2}}^{\log}(m_{2}) = \max \Big\{P_{k}^{\log}+I_{L_{1}\rightarrow R_{k}}^{\log}(m_{1})+I_{L_{3}\rightarrow R_{k}}^{\log}(m_{3})\Big\},
$}
\end{equation}
\begin{equation}
\label{eqn:log4}
\resizebox{.89\linewidth}{!}{$
I_{R_{k}\rightarrow L_{3}}^{\log}(m_{3}) = \max \Big\{P_{k}^{\log}+I_{L_{1}\rightarrow R_{k}}^{\log}(m_{1})+I_{L_{2}\rightarrow R_{k}}^{\log}(m_{2})\Big\},
$}
\end{equation}
\subsubsection{Layer Node Updating}
\begin{equation}
\label{eqn:log5}
I_{L_{j}\rightarrow R_{1}}^{\log}(m) = I_{R_{2}\rightarrow L_{j}}^{\log}(m),
\end{equation}
\begin{equation}
\label{eqn:log6}
I_{L_{j}\rightarrow R_{2}}^{\log}(m) = I_{R_{1}\rightarrow L_{j}}^{\log}(m),
\end{equation}
\subsubsection{Probability Calculating and Symbol Judging}
\begin{equation}
\label{eqn:81}
Q_{L_{j}}^{\log}(m) = I_{R_{1}\rightarrow L_{j}}^{\log}(m) + I_{R_{2}\rightarrow L_{j}}^{\log}(m).
\end{equation}

\subsection{Early Termination}
Early termination is based on the belief judgement for each layer node and resource node \cite{chen2005improved}.
Our judgement steps are:
\begin{enumerate}
  \item Create a zero-matrix to record the stability condition of beliefs, which denotes all the beliefs are unstable.
  \item Judge the stability of all beliefs per iteration. If $| \frac{V-V_\text{temp}}{V_\text{temp}} | \leq \epsilon,(\epsilon > 0)$, the beliefs are stable, and the corresponding value in the matrix is set as ``$1$''.
  \item When the stability matrix become a all-ones matrix, beliefs of all layer nodes and resource nodes are stable, and the convergence is achieved. Then, the iterative decoding terminates.
\end{enumerate}
Here, $V_\text{temp}$ and $V$ are the belief values in the previous and present iteration, respectively. $\epsilon$ is a judgment constant. The DMPA with early termination is shown in Alg. \ref{alg:PLA}. The Max-Log version is similar and omitted.
\begin{algorithm}[htbp]
\caption{DMPA with Early Termination}
\label{alg:PLA}
\begin{algorithmic}[1]
\Require
$\bf{y}$,
$I_{\mathrm{max}}$,
and $\epsilon$
\State
\textbf{Iteration:}
\State
\ \ \ \textbf{for} $t=1:I_{\mathrm{max}}$
\State
\ \ \ \ \ \ \ Set stability matrix $\bf{S} = \bf{0}$
\State
\ \ \ \ \ \ \ Update beliefs $V$
\State
\ \ \ \ \ \ \ \textbf{for} $j=1:N$\\
\ \ \ \ \ \ \ \ \ \ \ $temp = \big |V_{j}^{(t)}-V_{j}^{(t-1)} / V_{j}^{(t-1)} \big|$
\State
\ \ \ \ \ \ \ \ \ \ \ \textbf{if} $temp \leq \epsilon$\\
\ \ \ \ \ \ \ \ \ \ \ \ \ \ $S_{j} = 1$\\
\ \ \ \ \ \ \ \ \ \ \ \textbf{end if}\\
\ \ \ \ \ \ \ \textbf{end for}
\State
\ \ \ \ \ \ \ \textbf{if} $\bf{S} = \bf{1}$\\
\ \ \ \ \ \ \ \ \ \ \textbf{break}\\
\ \ \ \ \ \ \ \textbf{end if}\\
\ \ \ \textbf{end for}
\State
\textbf{Judgement$^{\text{early}}$:}
\State
\ \ \ Compute beliefs \\
\ \ \ Decide $\bf{\hat{u}}$
\Ensure
${\bf{\hat{u}}} = \{\hat{u}_{1}, \hat{u}_{2}, ..., \hat{u}_{6}\}$
\end{algorithmic}
\end{algorithm}

\subsection{Self-Adaption Algorithm}
Self-adaption \cite{Wu2010Adaptive,Savin2008Self} is also based on stability judgement. Compared to the one in early termination, the judgement in self-adaption requires an extra step between 2) and 3):

``Forecast and adjust the belief of next iteration based on the convergence trend. If $\frac{V-V_\text{temp}}{V_\text{temp}} \geq \epsilon$, $V \Leftarrow \alpha V~\text{with}~\alpha > 1$, since the convergence trend makes values larger. Otherwise, if $\frac{V-V_\text{temp}}{V_\text{temp}} \leq -\epsilon$, $V \Leftarrow \beta V ~\text{with}~\beta < 1$.''

Now the DMPA with self-adaption is shown in Alg. \ref{alg:PLA1}. The Max-Log version is omitted.
\begin{algorithm}[htbp]
\caption{DMPA with Self-Adaption}
\label{alg:PLA1}
\begin{algorithmic}[1]
\Require
$\bf{y}$,
$I_{\mathrm{max}}$,
and $\epsilon$
\State
\textbf{Iteration:}
\State
\ \ \ \textbf{for} $t=1:I_{\mathrm{max}}$
\State
\ \ \ \ \ \ \ Set stability matrix $\bf{S} = \bf{0}$
\State
\ \ \ \ \ \ \ Update beliefs $V$
\State
\ \ \ \ \ \ \ \textbf{for} $j=1:N$\\
\ \ \ \ \ \ \ \ \ \ \ $temp = V_{j}^{(t)}-V_{j}^{(t-1)} / V_{j}^{(t-1)}$
\State
\ \ \ \ \ \ \ \ \ \ \ \textbf{if} $temp \geq \epsilon$\\
\ \ \ \ \ \ \ \ \ \ \ \ \ \ $V_{j}^{(t)} \leftarrow \alpha \cdot V_{j}^{(t)}$
\State
\ \ \ \ \ \ \ \ \ \ \ \textbf{elseif} $temp \leq -\epsilon$\\
\ \ \ \ \ \ \ \ \ \ \ \ \ \ $V_{j}^{(t)} \leftarrow \beta \cdot V_{j}^{(t)}$\\
\ \ \ \ \ \ \ \ \ \ \ \textbf{else} \\
\ \ \ \ \ \ \ \ \ \ \ \ \ \ $S_{j} = 1$\\
\ \ \ \ \ \ \ \ \ \ \ \textbf{end if}\\
\ \ \ \ \ \ \ \textbf{end for}
\State
\ \ \ \ \ \ \ \textbf{if} $\bf{S} = \bf{1}$\\
\ \ \ \ \ \ \ \ \ \ \textbf{break}\\
\ \ \ \ \ \ \ \textbf{end if}\\
\ \ \ \textbf{end for}
\State
\textbf{Judgement$^{\text{adapt}}$:}
\State
\ \ \ Compute beliefs \\
\ \ \ Decide $\bf{\hat{u}}$
\Ensure
${\bf{\hat{u}}} = \{\hat{u}_{1}, \hat{u}_{2}, ..., \hat{u}_{6}\}$
\end{algorithmic}
\end{algorithm}

\subsection{Initial Noise Reduction}
``Distributed matrix'' $\mathbf{D}$ is to reduce random error, enhance accuracy of initial value \cite{forney1998modulation}, and speed up the convergence.
\begin{figure}[htbp]
\centering
\includegraphics[width=.85\linewidth]{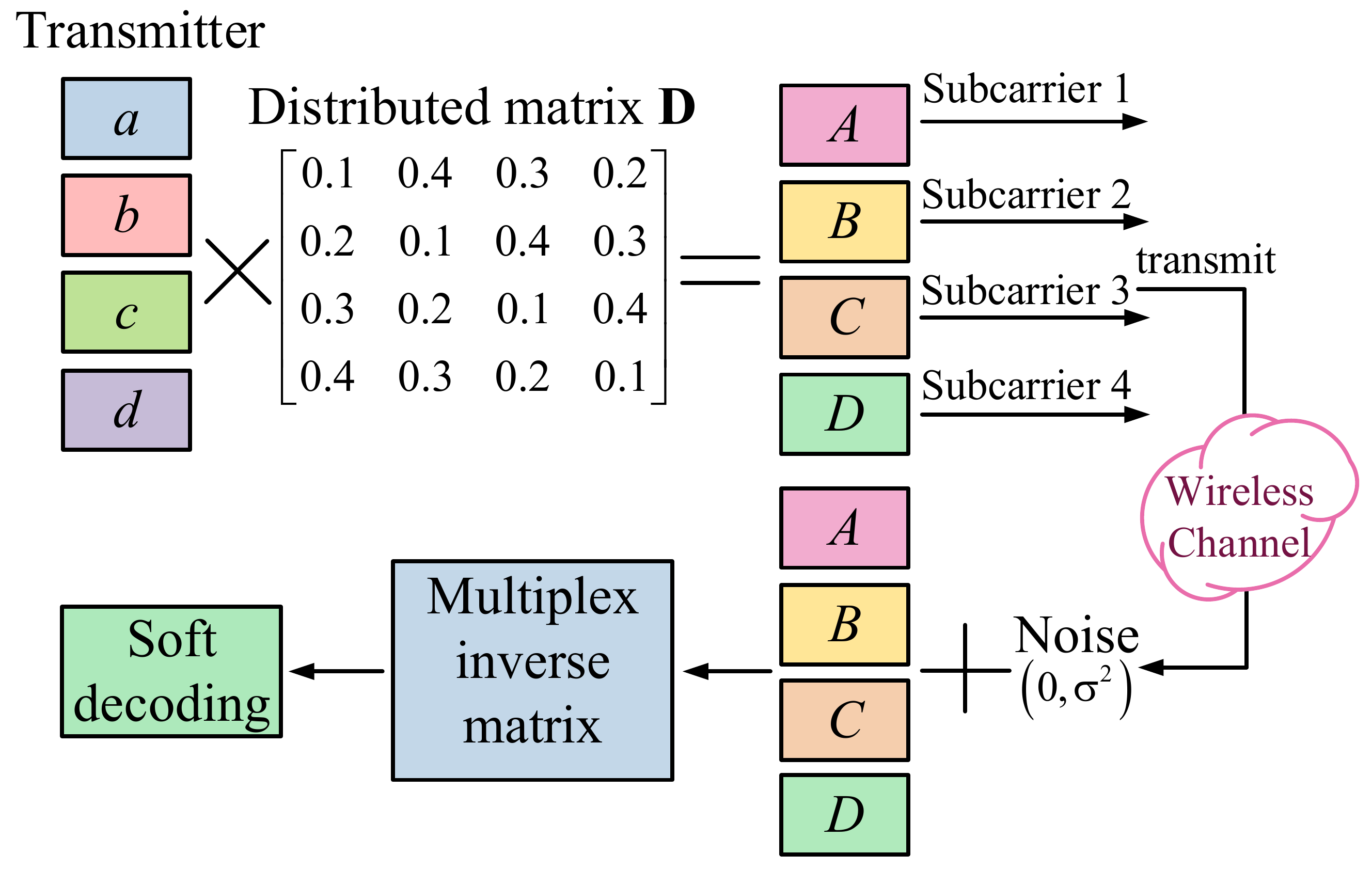}
\caption{\label{fig:noise_reduce}Procedure of initial noise reduction.}
\end{figure}

For the SCMA system in Fig. \ref{fig:noise_reduce}, we have the overlapped signals: $a$, $b$, $c$, and $d$ after multiplexing. Random error of these signals can be either positive or negative, which depends on the environment noise. Therefore, we can regroup signals and assign them to $4$ resource nodes. At the receiver, we can first recover the original signals according to the inverse of ``distributed matrix'' and then start the decoding. Compared to original transmitting scheme, each signal of specific resource node has a great chance to be added with both positive and negative random noises, which increases the accuracy of initial value. It is noted that $\mathbf{D}$ is not constant and can be adjusted according to the codebook and channel condition.

\subsection{Initial Probability Approximation}
Discussed above, the calculation of initial probability results in high computational complexity, which is obvious in Max-Log decoding. Thus, suitable approximations in \emph{Initialization} are expected to improve calculation efficiency and reduce latency with little performance loss. For SCMA decoding, the purpose of iterative updating is to find the symbol with the largest probability. Hence, the absolute value is not that critical to make a decision. We can still ensure the detection correctness even with relative beliefs. The relative magnitude is determined by the initial probability and the initial value of different users in \emph{Initialization}. Now, we carry out the approximation in steps: i) simplify the initial probability calculation by reducing operations with large complexity; ii) adjust the initial value of different users according to the relative magnitude determined by initial probabilities; iii) update beliefs iteratively based on the relative values. The formulae of initial probabilities in DMPA become:
\begin{equation}
\label{eqn:21}
\resizebox{.89\linewidth}{!}{$
P_{k}(y_{k}|x_{k,1},x_{k,2},x_{k,3},N_{0}) = e^{-\parallel y_{k}-(x_{k,1}+x_{k,2}+x_{k,3})\parallel^{2}/N_{0}},
$}
\end{equation}
For square and division, which are of higher complexity, DMPA approximations $1$ to $3$ are proposed:
\begin{equation}
\label{eqn:211}
\resizebox{.89\linewidth}{!}{$
P_{k}(y_{k}|x_{k,1},x_{k,2},x_{k,3},N_{0}) = e^{-\parallel y_{k}-(x_{k,1}+x_{k,2}+x_{k,3})\parallel/N_{0}},
$}
\end{equation}
\begin{equation}
\label{eqn:2111}
\resizebox{.89\linewidth}{!}{$
P_{k}(y_{k}|x_{k,1},x_{k,2},x_{k,3},N_{0}) = e^{-\parallel y_{k}-(x_{k,1}+x_{k,2}+x_{k,3})\parallel^{2}},
$}
\end{equation}
\begin{equation}
\label{eqn:21111}
\resizebox{.89\linewidth}{!}{$
P_{k}(y_{k}|x_{k,1},x_{k,2},x_{k,3},N_{0}) = e^{-\parallel y_{k}-(x_{k,1}+x_{k,2}+x_{k,3})\parallel},
$}
\end{equation}
Similarly, we have Max-Log approximations $1$ to $3$ as follows.
\begin{equation}
\label{equ:log11}
\resizebox{.89\linewidth}{!}{$
P_{k}^{\log}(y_{k}|x_{k,1},x_{k,2},x_{k,3},N_{0}) = -\frac{1}{N_{0}}\parallel y_{k}-(x_{k,1}+x_{k,2}+x_{k,3})\parallel,
$}
\end{equation}
\begin{equation}
\label{equ:log111}
\resizebox{.89\linewidth}{!}{$
P_{k}^{\log}(y_{k}|x_{k,1},x_{k,2},x_{k,3},N_{0}) = -\parallel y_{k}-(x_{k,1}+x_{k,2}+x_{k,3})\parallel^{2},
$}
\end{equation}
\begin{equation}
\label{equ:log1111}
\resizebox{.89\linewidth}{!}{$
P_{k}^{\log}(y_{k}|x_{k,1},x_{k,2},x_{k,3},N_{0}) = -\parallel y_{k}-(x_{k,1}+x_{k,2}+x_{k,3})\parallel,
$}
\end{equation}
Analysis below will show these approximations have different effects on error performance and computational complexity.

\section{Results and Analysis}\label{sec:Results}
\subsection{Error-Rate Performance}
\begin{figure*}[htbp]
  \centering
  \subfigure[Error-rate performance of DMPA with different approximation.]{
    \label{fig:DMPA_all} 
    \includegraphics[width=7in]{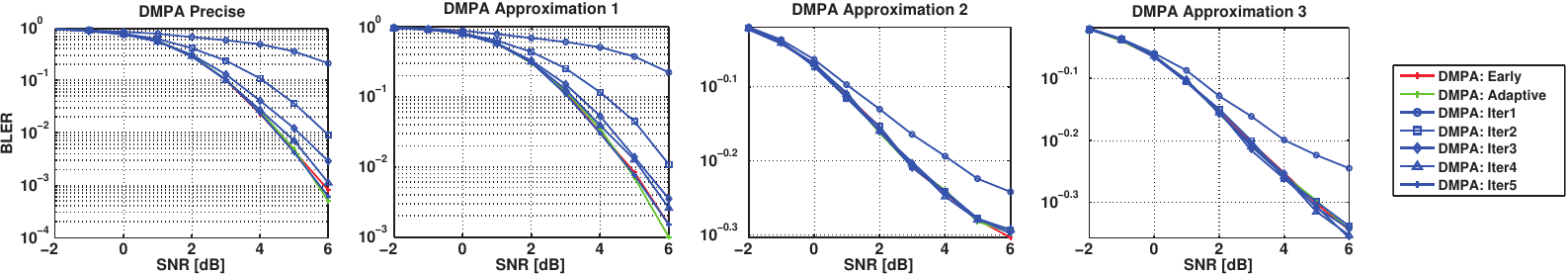}}
  \subfigure[Error-rate performance of Max-Log with different approximation.]{
    \label{fig:Max-Log_all} 
    \includegraphics[width=7in]{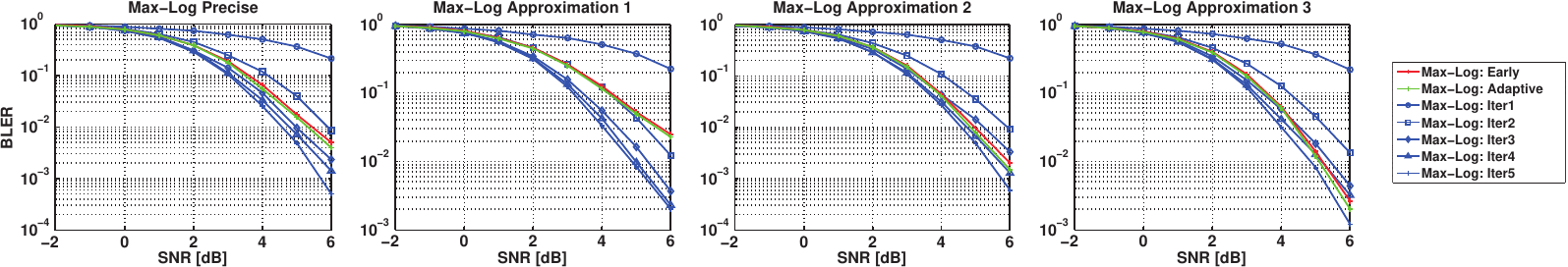}}
  \caption{Error-rate performance of SCMA with different detecting methods.}
  \label{fig:Error_performance_of_SCMA} 
\end{figure*}

The $6$-user SCMA system is simulated. Additional white Gaussian noise (AWGN) is assumed. The maximum iteration number is $5$. Results are give in Fig. \ref{fig:Error_performance_of_SCMA}.

Fig. \ref{fig:DMPA_all} shows the BLER performance of DMPA algorithm with different approximations, different iterations, early termination, self-adaption, and initial noise reduction. Fig. \ref{fig:Max-Log_all} shows the curves of Max-Log algorithm. According to Fig. \ref{fig:Error_performance_of_SCMA}, we see

\begin{enumerate}
  \item DMPA/Max-Log with more iterations enjoys better performance, but the improvement is limited when iteration number is sufficiently large. Shown by numerical results, DMPA/Max-Log with $3$ iterations is a good choice in real implementation.
  \item The average iteration number of early termination or adaptive scheme is around $3$, but the performance is similar DMPA with $5$ iterations. Results with different parameters reveal that self-adaption performs better in high SNR. Thus, the adjusting factor in self-adaption is supposed to be smaller at higher SNR.
  \item DMPA and Max-Log have similar performance without approximation. However, since DMPA heavily depends on $N_0$, approximations without precise $N_0$ will cause unbearable performance loss. On the other hand, Max-Log algorithm is not sensitive to $N_0$, and its approximations without exact $N_0$ can still achieve good performance. Therefore, Max-Log is preferred.
\end{enumerate}

Now, we figure out that suitable configurations for hardware implementation are: i) Max-Log approach; ii) $3$ iterations; iii) early termination and self-adaption; iv) Approximation $2$ or $3$, and v) initial noise reduction.

\subsection{Computational Complexity}
Suppose the symbol set size for each user is $M$, the number of physical resources is $N$, the user number is $K$, and the maximum iteration number is $I$. Then, we summarize the computational complexity of different decoding methods in Table \ref{tab:complexity}. Compared with other methods, the proposed method has the lowest computational complexity, while maintaining the error performance. In fact, the proposed method is similar to Max-Log, but has lower complexity in \emph{Initialization} due to the approximation. For a real system, $M$ and $N$ are usually large, the number of multiplications and divisions will makes other methods not suitable for implementation. However, as discussed above, the proposed algorithm is multiplication/division-free with Approximation $3$. Therefore, it can intensively improve the computational efficiency and reduce the latency, making it more applicable for hardware implementation in Section \ref{sec:VLSI}. The VLSI implementation results in Table \ref{tab:fpgare} will further verify that the proposed decoder's hardware efficiency over the SOA design.
\begin{table*}[htbp]
\centering
\renewcommand\arraystretch{1.2}
\footnotesize
\caption{\label{tab:complexity}Comparison of Computational Complexity for Different Decoding Algorithms}
\begin{tabular}{clllll}
\Xhline{1.0pt}
  \multirow{2}{*}{Procedure} & \multirow{2}{*}{Operation} & \multirow{2}{*}{This work} & DMPA \cite{SCMA_archi} & Max-Log \cite{Max-log} & Pruned DMPA \cite{tree_2} \\
  & & & [APCCAS '17] & [China Comm. Dec. '15] & [DSP '16]\\
  \Xhline{.5pt}
\multirow{3}{*}{\minitab[c]{Initial probability \\ calculation}} & \cellcolor{gray!15}ADD & \cellcolor{gray!15}$2M^{3}N/T_{\text{adp}}$ & \cellcolor{gray!15}$3M^{3}N/T_{\text{MPA}}$ & \cellcolor{gray!15}$3M^{3}N/T_{\text{Max-Log}}$ & \cellcolor{gray!15}$3M^{3}N/T_{\text{tree}}$ \\
& MUL & $0$ & $3M^{3}N/T_{\text{MPA}}$ & $3M^{3}N/T_{\text{Max-Log}}$ & $3M^{3}N/T_{\text{tree}}$ \\
& \cellcolor{gray!15}EXP & \cellcolor{gray!15}$0$ & \cellcolor{gray!15}$M^{3}N/T_{\text{MPA}}$ & \cellcolor{gray!15}0 & \cellcolor{gray!15}$M^{3}N/T_{\text{tree}}$ \\ \Xhline{.5pt}
\multirow{3}{*}{\minitab[c]{Resource node \\ updating}} & ADD & $2 \cdot 3M^{3}N$ & $3M^{3}N$ & $2 \cdot 3M^{3}N$ & $3M^{3}N$ \\
& \cellcolor{gray!15}MUL & \cellcolor{gray!15}$0$ & \cellcolor{gray!15}$2 \cdot 3M^{3}N$ & \cellcolor{gray!15}0 & \cellcolor{gray!15}$2 \cdot 3M^{3}N$ \\
& MAX & $3M^{3}N$ & 0 & $3M^{3}N$ & 0\\ \Xhline{.5pt}
\multirow{3}{*}{\minitab[c]{Layer node \\ updating}} & \cellcolor{gray!15}ADD & \cellcolor{gray!15}0 & \cellcolor{gray!15}$2MK$ & \cellcolor{gray!15}0 & \cellcolor{gray!15}$2MK$\\
& MUL & 0 & $2MK$ & 0 & $2MK$ \\
& \cellcolor{gray!15}SWOP & \cellcolor{gray!15}$2MK$ & \cellcolor{gray!15}$2MK$ & \cellcolor{gray!15}$2MK$ & \cellcolor{gray!15}$2MK$ \\ \Xhline{.5pt}
\multirow{3}{*}{\minitab[c]{Users' symbol \\ judgement}} & ADD & $MK$ & 0 & $MK$ & 0 \\
& \cellcolor{gray!15}MUL & \cellcolor{gray!15}0 & \cellcolor{gray!15}$MK$ & \cellcolor{gray!15}0 & \cellcolor{gray!15}$MK$ \\
& MAX & 0 & $MK$ & 0 & $MK$ \\
\Xhline{1.0pt}
\end{tabular}
\end{table*}

\subsection{Performance/Complexity Trade-Off Analysis}
\begin{figure}[htbp]
\centering
\includegraphics[width=.75\linewidth]{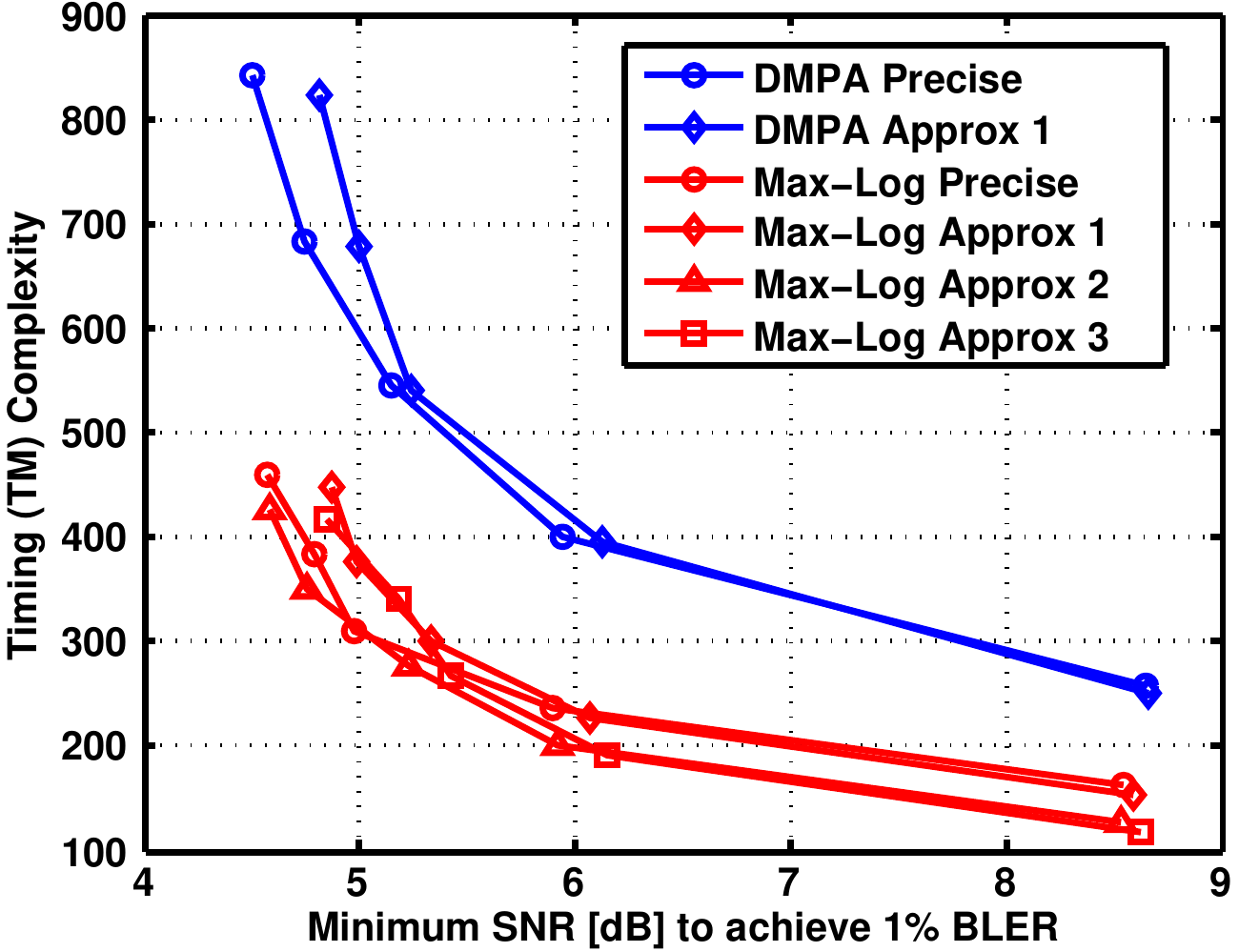}
\caption{Performance/complexity trade-off analysis of DMPA algorithm.}
\label{fig:tradeoff}
\end{figure}

Fig. \ref{fig:tradeoff} illustrates the trade-off between error performance and computational complexity of proposed methods. The minimum required SNR to achieve $1\%$ BER is employed as a metric. The complexity is given by Timing (TM) complexity, which is in term of iteration number. Fig. \ref{fig:tradeoff} shows the trade-off of DMPA with approximations. It is clear that Max-Log with Approximation $3$ provides the best performance/complexity trade-off.



\section{Hardware Architecture}\label{sec:Hardware}
The hardware architecture of the Max-Log DMPA is discussed. Timing optimization and folding technique are introduced for higer efficiency.

\subsection{Overall Architecture}
\begin{figure}[htbp]
\centering
\includegraphics[width=.8\linewidth]{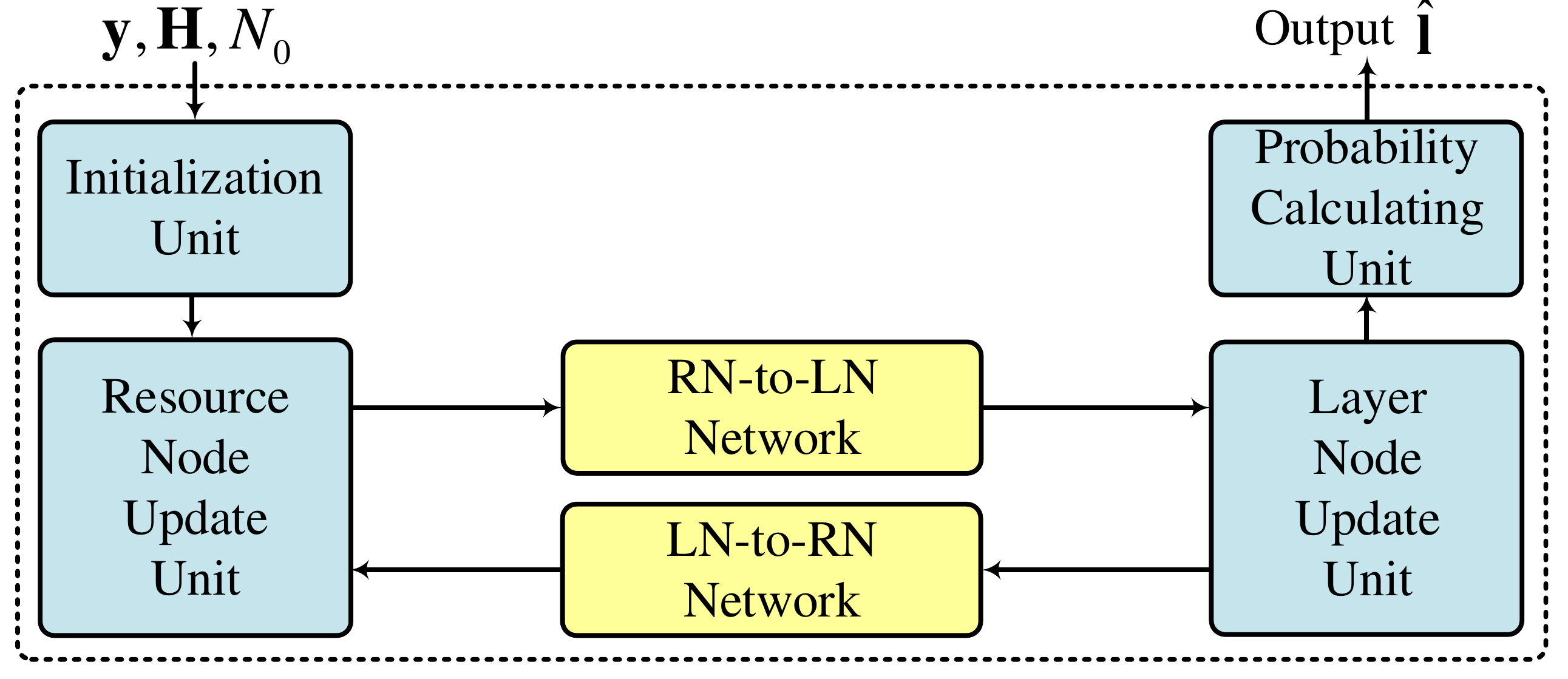}
\caption{Overall architecture of DMPA.}
\label{fig:General architecture of DMPA}
\end{figure}
The overall architecture is shown in Fig. \ref{fig:General architecture of DMPA}. It has $4$ units and $2$ memory networks, which are RN-to-LN and LN-to-RN networks for $I_{R\rightarrow L}$ and $I_{L\rightarrow R}$, respectively. The elementary units are \emph{Initialization Unit}, \emph{Resource Node Update Unit}, \emph{Layer Node Update Unit}, and \emph{Probability Calculating Unit}, which execute steps indicated by Eq. (\ref{equ:log1}) to Eq. (\ref{eqn:81}), respectively. The iterative calculation is done by \emph{Resource Node Update Unit} and \emph{Layer Node Update Unit}, both of which could not start current propagation unless all previous data have been calculated. We call this data updating interval a ``step". Optimization details of this scheduling will be discussed below.

\subsection{Stage-Level Scheduling Optimization}
\begin{figure}[htbp]
\centering
\includegraphics[width=\linewidth]{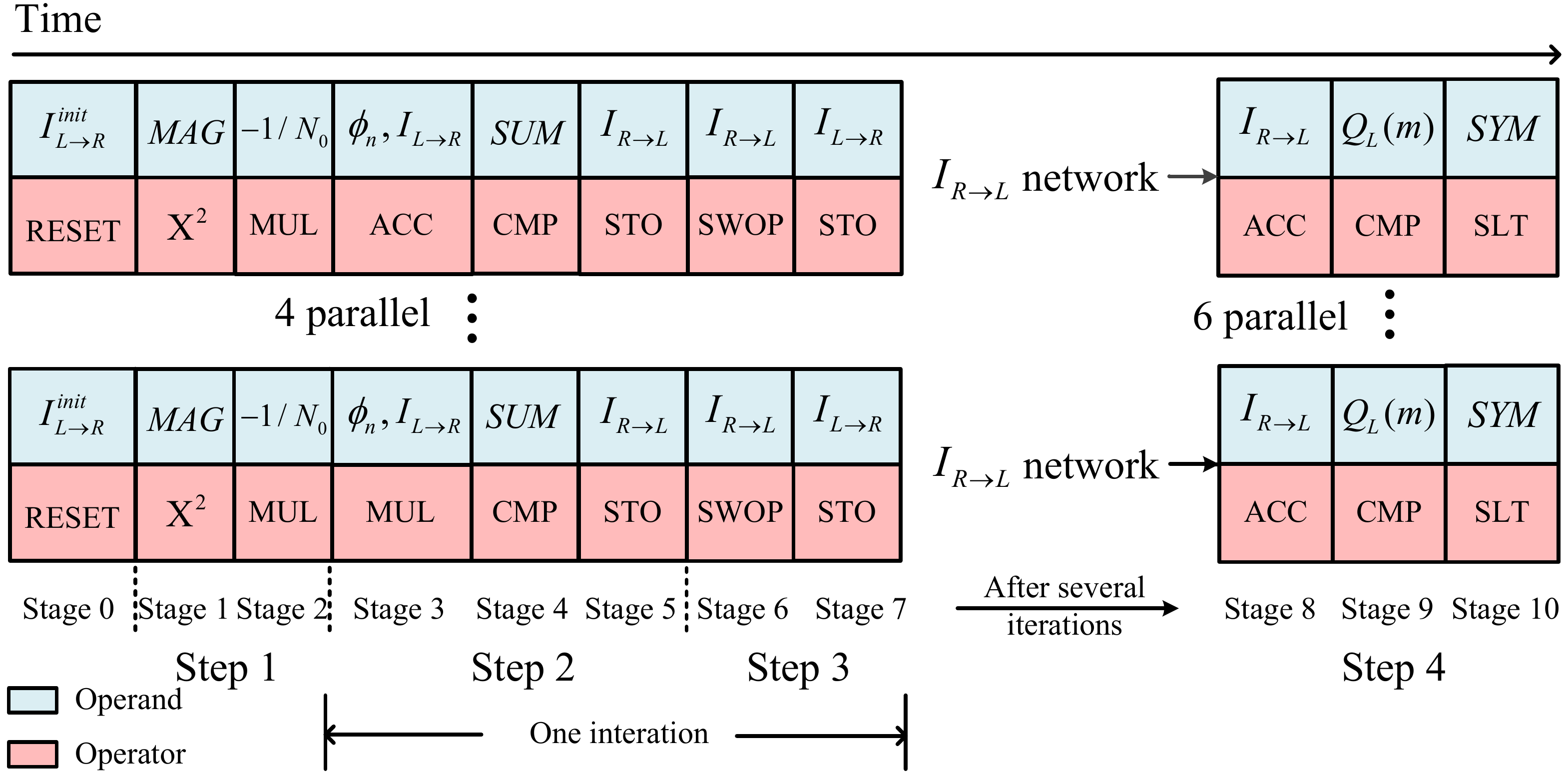}
\caption{Stage-level scheduling.}
\label{fig:Scheduling of stage-level}
\end{figure}
The proposed stage-level scheduling is a finer-grained optimization over the step-level scheduling. With this stage-level scheduling, it is convenient to insert deep pipelines to achieve a higher throughput \cite{simon1994new,monteiro1996retiming,parhi1992synthesis}. Compared with step-level scheduling, updating of stage-level scheduling does not have to wait for the completion of data computation from the previous unit, which therefore avoids low hardware-efficiency and long processing-latency. In sum, stage-level scheduling enjoys faster processing speed and higher hardware efficiency than the step-level one. Fig. \ref{fig:Scheduling of stage-level} shows the stage-level scheduling. It details each computing step to achieve a deeper pipelined structure.

%

\subsection{Folding}
\begin{figure}[htbp]
\centering
\includegraphics[width=.85\linewidth]{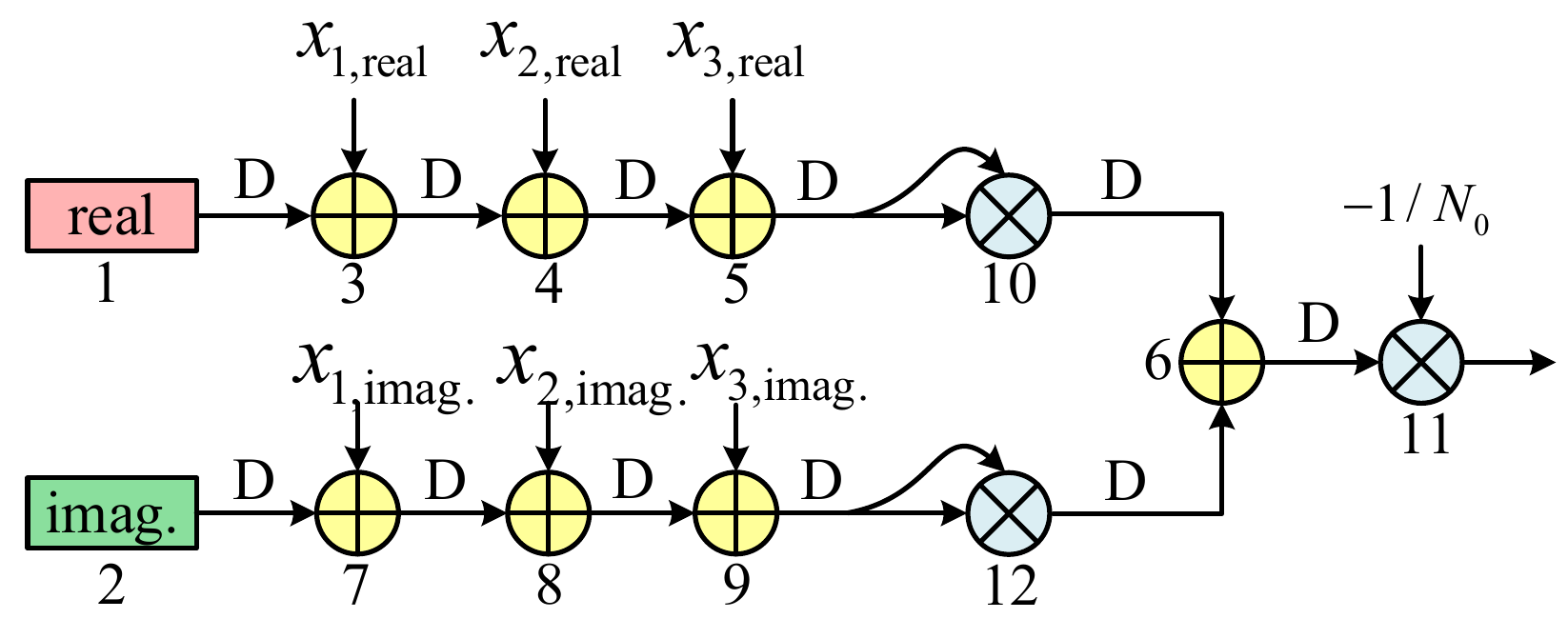}
\caption{Original hardware of Step $1$ before folding.}
\label{fig:Original hardware of step1 in need of folding}
\end{figure}

The architecture of stage-level DMPA turns out to be very complicated in form of data factor graph (DFG). To achieve an efficient architecture, folding technique is employed for further optimization. Since folding operation based on fine-grained architecture is difficult to be carried out, a folding scheme based on unit is considered. Fig.s \ref{fig:DFG of step-level architecture} and \ref{fig:DFG of stage-level architecture} in appendix shows the entire step- and stage-level algorithms, respectively. Due to the page constraint, we only take a branch of \emph{Initialization Unit}, which is fully-paralleled in DFG, as an example to show proposed folding details. Folding transform of other units can be conducted in the similar fashion. The DFG of the branch in \emph{Initialization Unit} is shown in Fig. \ref{fig:Original hardware of step1 in need of folding}.

The folding includes $3$ steps: i) construct folding sets and folding equations, ii) analysis life span, and iii) allocate registers. More details of this method are explained by \cite{parhi2007vlsi}.

\subsubsection{Folding Sets and Folding Equations}
Set the folding factor to $7$, we can obtain the following folding sets:
\begin{equation}\label{eq:s}
\left\{
\begin{aligned}
  S_{in}&= \{1,2,\phi,\phi,\phi,\phi,\phi\},\\
  S_{A} &= \{3,4,5,6,7,8,9\},\\
  S_{M} &= \{10,11,12,\phi,\phi,\phi,\phi\},
\end{aligned}
\right.
\end{equation}
where $S_{in}$, $S_{A}$, and $S_{M}$ denote the folding sets for inputs, adders, and multipliers, respectively.

Then, folding equations can be derived based on the given folding sets
{\small
\begin{equation}\label{eq:d}
\left\{
\begin{alignedat}{3}
D_{F}(1\rightarrow3)&= 0,D_{F}(3\rightarrow4)&&= 7,D_{F}(4\rightarrow5) &&= 7,\\
D_{F}(2\rightarrow7)&= 3,D_{F}(7\rightarrow8)&&= 7,D_{F}(8\rightarrow9) &&= 7,\\
D_{F}(5\rightarrow10)&= 4,D_{F}(10\rightarrow6)&&= 8,D_{F}(6\rightarrow11) &&= 4,\\
D_{F}(9\rightarrow12)&= 2,D_{F}(12\rightarrow6)&&= 6,
\end{alignedat}
\right.
\end{equation}}
where $D_{F}(x\rightarrow y)$ denotes the number of delays on the path from $x$ to $y$.

\subsubsection{Life Time Analysis}
\begin{figure}[htbp]
\centering
\includegraphics[width=7 cm]{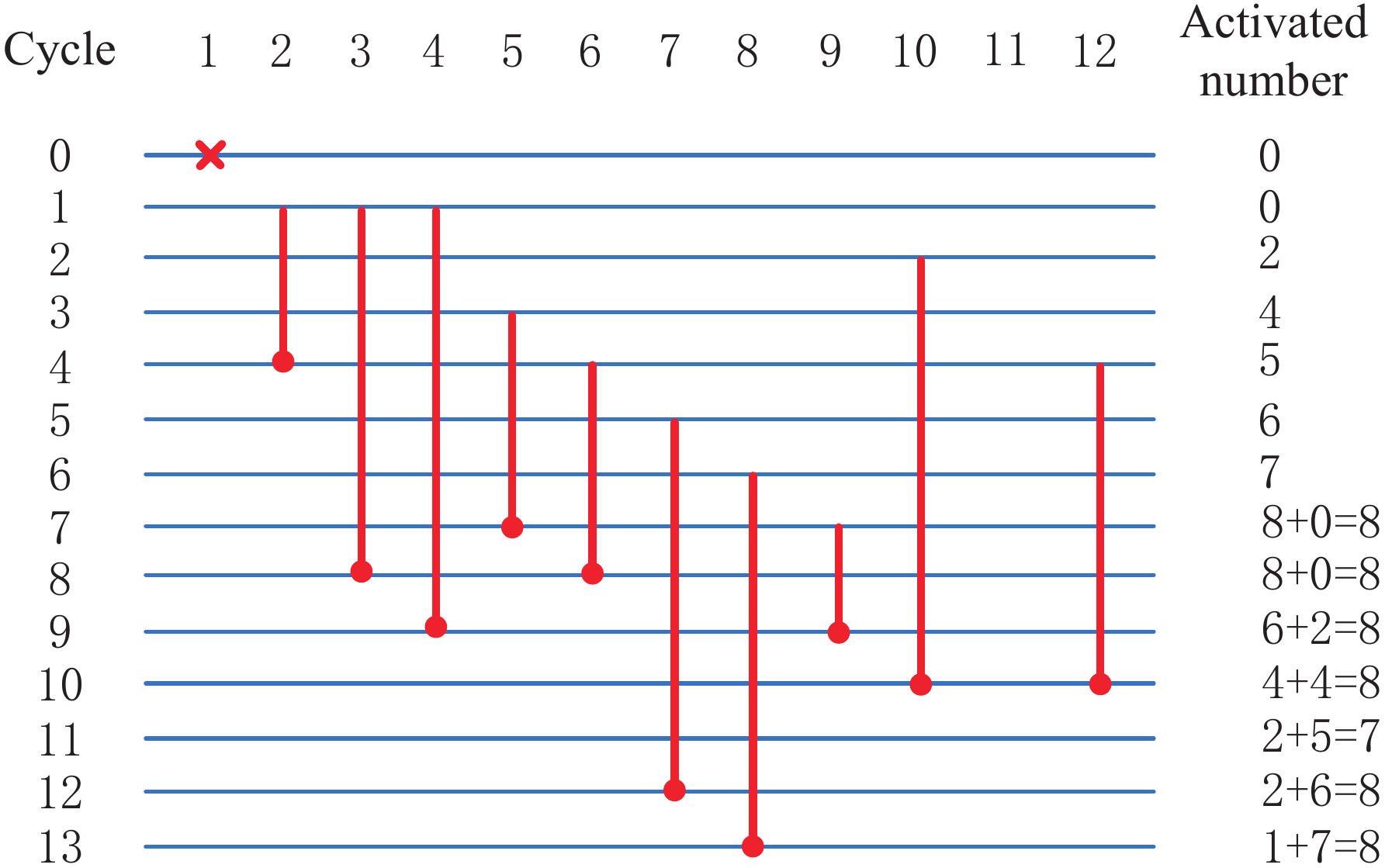}
\caption{Life time figure.}
\label{fig:Life time figure}
\end{figure}

Life span analysis is demonstrated in the form of life time figure as shown in Fig. \ref{fig:Life time figure}. It is achieved from folding equations. One thick line in the figure represents survival time of certain data. Activated number shows number of data in use at the moment \cite{parhi1994calculation}. According to Fig. \ref{fig:Life time figure}, we see that this folding architecture requires at least $8$ registers.

\subsubsection{Register Allocation}
\begin{figure}[htbp]
\centering
\includegraphics[width= 8cm]{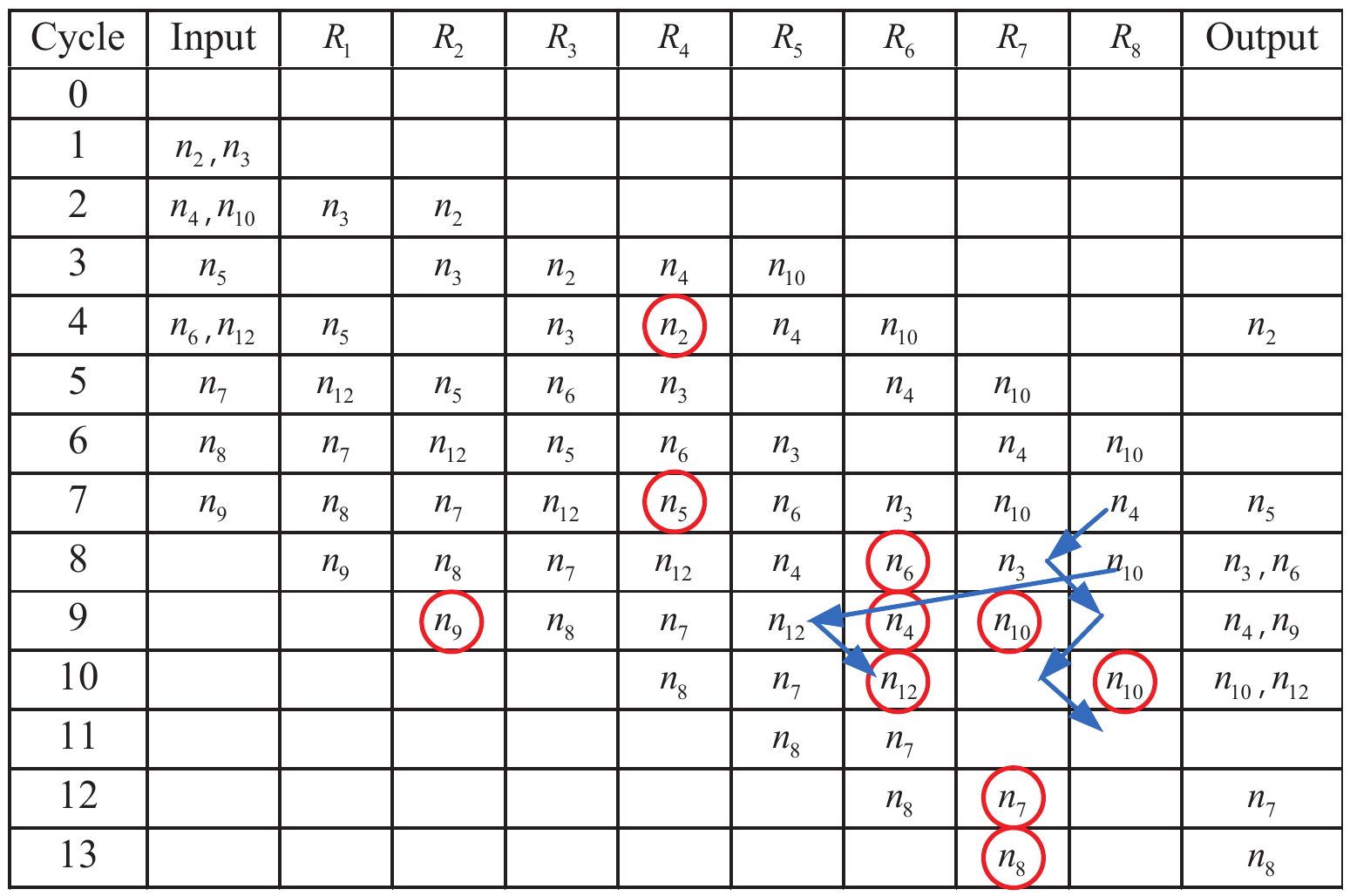}
\caption{Register allocation table of folding architecture.}
\label{fig:Register allocation table of folding architecture}
\end{figure}

The forward-backward scheme of register allocation is employed based on life span analysis \cite{parhi1992systematic}. The specific allocation process is displayed in Fig. \ref{fig:Register allocation table of folding architecture}.

After all the steps, we can finally obtain the folded architecture of the branch in \emph{Initialization Unit}.

\subsection{Hardware Architecture and Loop Analysis}
\begin{figure}[htbp]
\centering
\includegraphics[width=\linewidth]{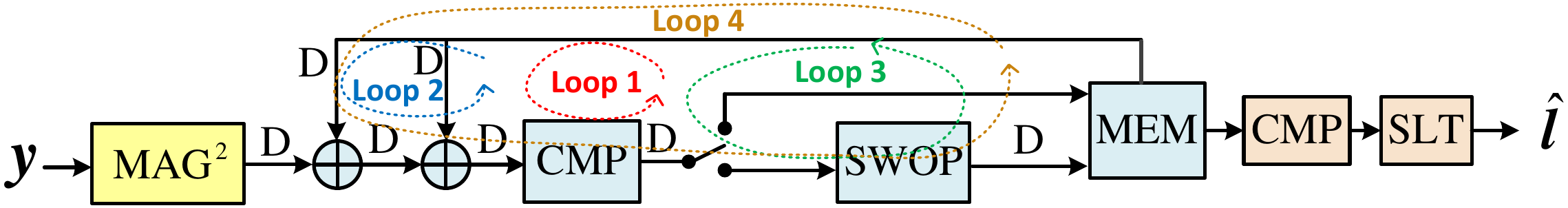}
\caption{Hardware architecture of DMPA.}
\label{fig:Final architecture of DMPA}
\end{figure}

The final stage-level folded architecture of DMPA, which is illustrated at module-level in Fig. \ref{fig:Final architecture of DMPA}.
Lower hardware cost and reasonable processing speed become its main advantages.

\begin{table}[htbp]
\centering
\renewcommand\arraystretch{1.2}
\footnotesize
\caption{Cost of Different Architectures for ``$J=6$''}
\begin{tabular}{llcc}
\Xhline{1.0pt}
\multirow{2}{*}{Different architectures} &\multirow{2}{*}{Cycles} &  \multicolumn{2}{c}{Hardware cost (main untis)}\\ \cline{3-4}
&   & Adders (Comparators)  & Multipliers\\
\Xhline{0.5pt}
\rowcolor{gray!15}Original & $80$ & $52$ & $12$\\
Stage-level folded & $300$ & $4$ & $2$\\
\Xhline{1.0pt}
\end{tabular}
\end{table}

The loop bound analysis \cite{chao1993iteration,ito1995determining} of this folded architecture is also given here. Suppose the processing time of an adder, a comparator, and a swopper are $T_{A}, T_{C}$, and $T_{S}$, respectively.
We can obtain the results listed by Table \ref{tab:loop bound analysis}.
\begin{table}[htbp]
\centering
\renewcommand\arraystretch{1.2}
\footnotesize
 \caption{Loop Bound Analysis}
 \label{tab:loop bound analysis}
 \begin{tabular}{cccccl}
\Xhline{1.0pt}
  Loop & ADD & CMP & SWOP & Delay & Loop bound\\
\Xhline{.5pt}
  \rowcolor{gray!15}$1$ & $1$ & $1$ & $0$ & $3$ & $(T_{A}+T_{C})/3$\\
  $2$ & $2$ & $1$ & $0$ & $4$ & $(2T_{A}+T_{C})/4$\\
  \rowcolor{gray!15}$3$ & $1$ & $1$ & $1$ & $4$ & $(T_{A}+T_{C}+T_{S})/4$\\
  $4$ & $2$ & $1$ & $1$ & $5$ & $(2T_{A}+T_{C}+T_{S})/5$\\
\Xhline{1.0pt}
 \end{tabular}
\end{table}

Thus, the iteration bound is calculated as follows:
\begin{equation}
\resizebox{.89\linewidth}{!}{$
T_{\infty}\!=\!
\max
\left\{
\begin{alignedat}{3}
&{\textstyle\frac{T_{A}+T_{C}}{3}},&&{\textstyle\frac{2T_{A}+T_{C}}{4}},{\textstyle\frac{T_{A}+T_{C}+T_{S}}{4}},&&{\textstyle\frac{2T_{A}+T_{C}+T_{S}}{5}}\\
\end{alignedat}
\right\}
$}
\end{equation}

%

\section{VLSI Implementation}\label{sec:VLSI}
The proposed decoder's VLSI implementation is given and compared to two SOA baselines. The first is the DMPA decoder \cite{SCMA_archi}, and the second is the SMPA decoder \cite{SMPA_2}. As both baselines do not consider folding, the proposed decoder does not either for fair comparison. But if all designs are folded, the proposed decoder's advantages remain. Discussed previously, the proposed decoder is based on: i) Max-Log approach; ii) early termination and self-adaption; iii) Approximation $3$, and iv) initial noise reduction. Since the SMPA decoder employed $5$ iterations, $1$ up to $5$ iterations are considered, though $3$ turns out to be efficient per our analysis. Both the proposed decoder and DMPA decoder are implemented with Xilinx Virtex-7 XC7VX690T FPGA. The results of SMPA decoder is scribed from \cite{SMPA_2}, since it is implemented with ASIC. The frequency is $500$ MHz. The input quantization is $8$-bit for both real and imaginary parts, and the intermediate quantization is $16$.

\subsection{Module Details of Proposed Decoder}\label{sec:sub:details}
The proposed decoder consists of four basic parts as shown in Fig. \ref{fig:General architecture of DMPA}: initialization module, layer node updating network, resource node updating network, and symbol judging module. The design details are presented as follows.

\subsubsection{Initialization Module}\label{sec:subsub:IM}
It calculates initial belief of each user with the received signal and inner codebook. The received signal is made up of $4$ complex resource nodes, thus the input is $8$-parallel. Each of them has the quantization length of $8$. It is noted that the output belief has the quantization length of $16$, due to multiplication. The codebook is restored in memories, which costs $96$ memory blocks of $8$-bit length each.

\subsubsection{Resource Node Updating Network}\label{sec:subsub:rnun}
It calculates the sum of belief and outputs the largest, based on the approximated Jacobi's formula. It is made up of resource node updating units, where the input data are initial beliefs and layer node beliefs, and the output data are the $4$ resource node beliefs. The largest value is selected from $16$ intermediate beliefs, in $3$ steps of comparison with $14$ buffers. Thus, $56$ buffers are required by each unit, and $672$ by the entire network.

\subsubsection{Layer Node Updating Network}\label{sec:subsub:lnun}
It is made up of layer node updating units,
which normalize the input value and swop it by the inner connection. In each unit, the input data are resource node beliefs only, and the output data are the corresponding $4$ layer node beliefs. Four $16$-bit dividers are required per unit with $28$ clocks' delay. Hence, the whole network needs $48$ dividers. Besides, layer node beliefs would also be reset at the start of each frame of the received signals in layer node updating network.

\subsubsection{Symbol Judging Module}\label{sec:subsub:sjm}
It finds the largest belief and maps it to original source code according to the codebook of each user. Also, this module consists of $6$ smaller judging units, which perform the basic function for each user. In each unit, $4$ beliefs are compared with each other. Thus $2$ steps of comparison and $3$ buffers are required. Then, the entire module needs $18$ buffers.

The implementation comparison with the DMPA decoder is listed in Table \ref{tab:fpgare}. It shows the proposed decoder's advantages in both complexity and throughput, thanks to the log-domain processing and approximation approaches.
\begin{table}[htbp]
\tabcolsep 1mm
\renewcommand{\arraystretch}{1.2}
\centering
\footnotesize
\caption{FPGA Results for Different Decoders with $J/K=6/4$}
\label{tab:fpgare}
\begin{tabular}{lll}
\Xhline{1.0pt}
SCMA decoders & DMPA decoder \cite{SCMA_archi} & This work\\
\hline
\rowcolor{gray!15}
LUTs   & $139,205$ $(36\%)$   &  $82,909$ $(19\%)$ \\
Registers & $248,217$ $(28\%)$ & $109,997$ $(12\%)$  \\
\rowcolor{gray!15}
LUT-FF pairs	& $103,127$ $(36\%)$    & $52,203$ $(18\%)$ \\
DSP48E1s & $436$ $(12\%)$ & $436$ $(12\%)$\\
\rowcolor{gray!15}
Maximum frequency & $167.6$ MHz  & $359.1$ MHz \\
\Xhline{1.0pt}
\end{tabular}
\end{table}

Since speed is the main focus of our design, comparison results of throughput and latency with baselines are shown in Table \ref{tab:L_T}, where ``L'' for latency and ``T'' for throughput. As we can see from the table, the proposed SCMA decoder outperforms the SOA in both throughput and latency, and also meets the multi-Gbps and millisecond requirements of 3GPP. Though, SMPA decoder has complexity advantage, the proposed decoder's complexity can be further reduced with folding techniques.
\begin{table*}[htbp]
\centering
\renewcommand\arraystretch{1.2}
\footnotesize
\caption{\label{tab:L_T}Latency (L) in [$\mu$s] and Throughput (T) in [mb/s] for Different Decoders (Frequency: $500$ MHz)}
\begin{tabular}{lllllll}
\Xhline{1.0pt}
\rowcolor{gray!15}
User \# ($J$) & $6$ & $12$ & $24$ & $48$ & $96$ & $192$\\
Resource \# ($K$) & $4$ & $8$ & $16$ & $32$ & $64$ & $128$\\
\rowcolor{gray!15}
Iteration \# ($I_{\mathrm{max}}$) & L / T & L / T & L / T & L / T & L / T & L / T\\
\Xhline{.5pt}
\multicolumn{7}{c}{This work}\\
\Xhline{.5pt}
\rowcolor{gray!15}$1$ & $3.50$ / $857.14$ & $3.70$ / $1628.57$ & $3.96$ / $3012.85$ & $4.26$ / $5423.13$ & $4.58$ / $9490.48$ & $4.95$ / $16133.81$\\
$2$ & $5.50$ / $547.69$ & $5.84$ / $1040.62$ & $6.22$ / $1925.14$ & $6.66$ / $3465.25$ & $7.16$ / $6064.20$ & $7.74$ / $10309.14$\\
\rowcolor{gray!15}$3$ & $8.62$ / $349.96$ & $9.14$ / $664.93$ & $9.74$ / $1230.12$ & $10.42$ / $2208.46$ & $11.22$ / $3874.88$ & $12.12$ / $6587.30$\\
$4$ & $13.48$ / $223.62$ & $14.28$/ $424.88$ & $15.22$ / $786.02$ & $16.32$ / $1411.16$ & $17.56$ / $2475.96$ & $18.98$ / $4209.14$\\
\rowcolor{gray!15}$5$ & $21.00$ / $142.86$ & $22.12$ / $271.43$ & $23.42$ / $502.15$ & $24.94$ / $903.88$ & $26.68$ / $1581.78$ & $28.62$ / $2689.03$\\
\Xhline{.5pt}
\multicolumn{7}{c}{C. Yang \cite{SCMA_archi}, [DMPA decoder, APCCAS '17]} \\ \hline
\rowcolor{gray!15}$1$ & $5.60$ / $150.00$ & $5.80$ / $285.00$ & $6.06$ / $527.25$ & $6.36$ / $949.05$ & $6.78$ / $1660.84$ & $7.16$ / $2823.42$\\
$2$ & $10.92$ / $76.92$ & $11.26$ / $146.148$ & $11.64$ / $270.37$ & $12.08$ / $486.67$ & $12.58$ / $851.68$ & $13.16$ / $1447.85$\\
\rowcolor{gray!15}$3$ & $16.24$ / $51.72$ & $16.76$ / $98.27$ & $17.36$ / $181.80$ & $18.04$ / $327.23$ & $18.84$ / $572.66$ & $19.74$ / $973.52$\\
$4$ & $21.56$ / $38.96$ & $22.36$ / $74.02$ & $23.30$ / $136.94$ & $24.40$ / $246.50$ & $25.64$ / $431.37$ & $27.06$ / $733.34$\\
\rowcolor{gray!15}$5$ & $26.88$ / $31.25$ & $28.00$ / $59.38$ & $29.30$ / $109.84$ & $30.82$ / $197.72$ & $32.56$ / $346.01$ & $34.50$ / $588.21$\\
\Xhline{.5pt}
\multicolumn{7}{c}{K. Han \cite{SMPA_2}, [SMPA decoder, TCAS-I Oct. '17]}\\
\Xhline{.5pt}
$5$ & n.a. / $57$ & n.a. / n.a. & n.a. / n.a. & n.a. / n.a. & n.a. / $640$  & n.a. / n.a. \\
\Xhline{1.0pt}
\end{tabular}
\end{table*}

\section{Conclusion}\label{sec:Conclusion}
In this paper, simplifications such as log-domain calculation and probability approximation have been introduced to lower the complexity of SCMA's DMPA decoder. Early termination, adaptive decoding, and initial noise reduction are also proposed for faster convergence and better performance.
Hardware optimizations with folding and retiming are introduced. VLSI implementation results have confirmed the advantages of the proposed SCMA decoder for high-speed applications over the SOA designs. Future research will be directed towards further improvements on both algorithm and implementation.

\footnotesize
\bibliographystyle{IEEEtran}
\bibliography{IEEEabrv,mybib}

\begin{figure*}[htbp]
\centering
\includegraphics[width=\linewidth]{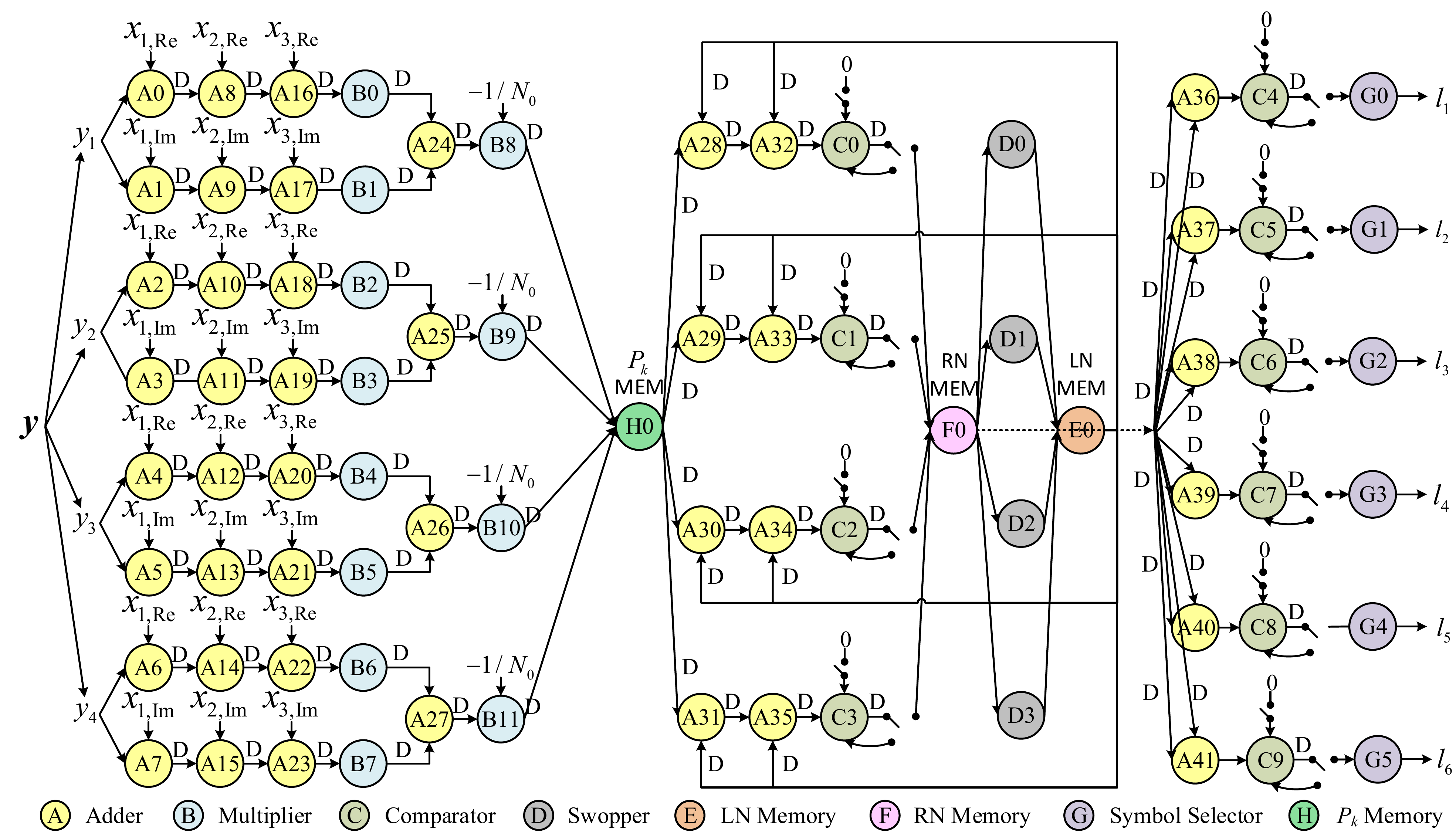}
\caption{Data flow graph (DFG) of step-level architecture.}
\label{fig:DFG of step-level architecture}
\end{figure*}

\begin{figure*}[htbp]
\centering
\includegraphics[width=\linewidth]{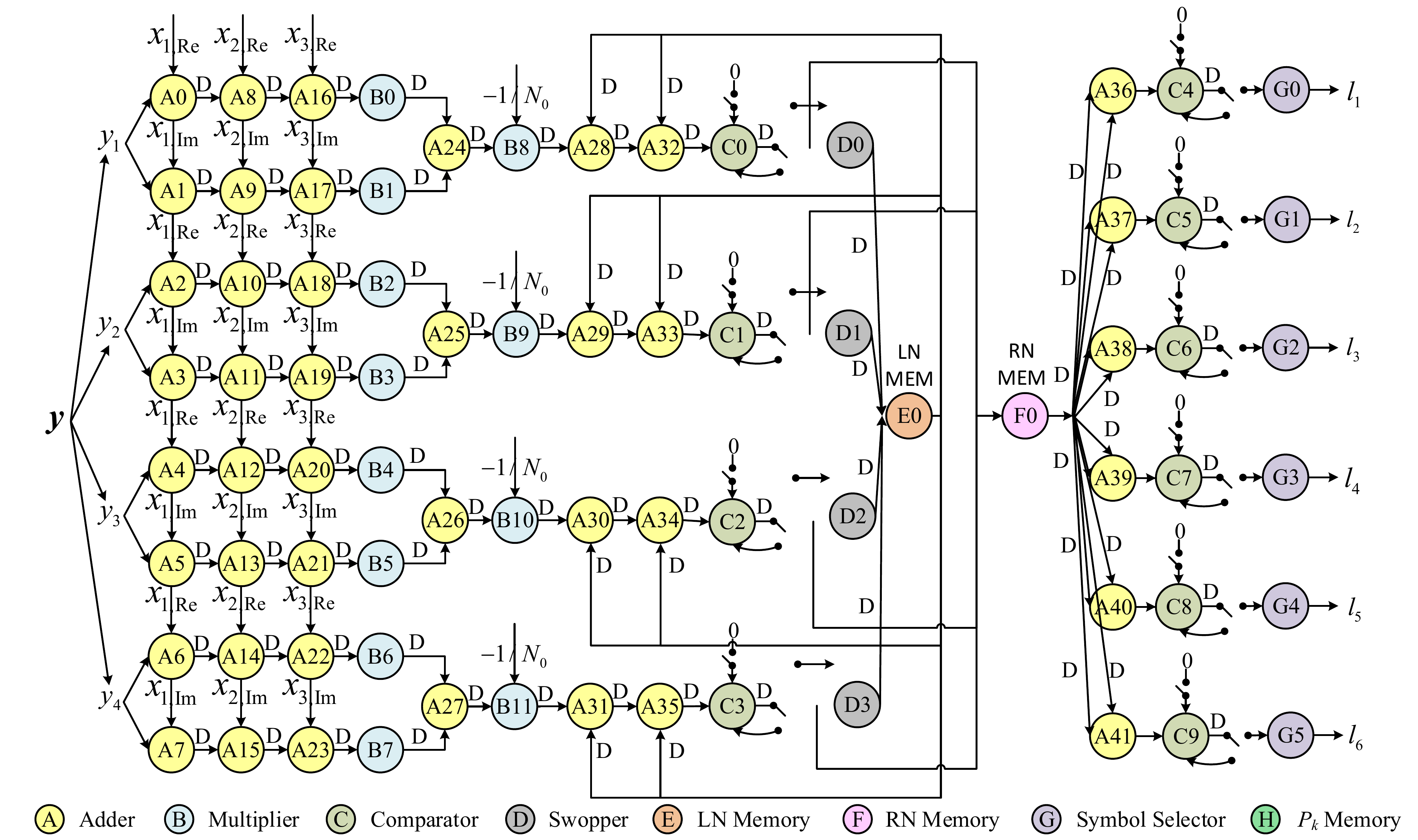}
\caption{Data flow graph (DFG) of stage-level architecture.}
\label{fig:DFG of stage-level architecture}
\end{figure*}

\end{document}